# Limitations of Agile Software Processes

**Dan Turk, Robert France**
Colorado State University
Fort Collins, Colorado, USA
dan.turk@colostate.edu
france@cs.colostate.edu

**Bernhard Rumpe**
Software & Systems Engineering
Munich University of Technology
Munich, Germany
Bernhard.Rumpe@in.tum.de

## ABSTRACT
Software developers and project managers are struggling to assess the appropriateness of agile processes to their development environments. This paper identifies limitations that apply to many of the published agile processes in terms of the types of projects in which their application may be problematic.

## INTRODUCTION
As more organizations seek to gain competitive advantage through timely deployment of Internet-based services, developers are under increasing pressure to produce new or enhanced implementations quickly [2,8]. Agile software development processes were developed primarily to address this problem, that is, the problem of developing software in "Internet time". Agile approaches utilize technical and managerial processes that continuously adapt and adjust to (1) changes derived from experiences gained during development, (2) changes in software requirements and (3) changes in the development environment.

Agile processes are intended to support early and quick production of working code. This is accomplished by structuring the development process into iterations, where an iteration focuses on delivering working code and other artifacts that provide value to the customer and, secondarily, to the project. Agile process proponents and critics often emphasize the code focus of these processes. Proponents often argue that code is the only deliverable that matters, and marginalize the role of analysis and design models and documentation in software creation and evolution. Agile process critics point out that the emphasis on code can lead to corporate memory loss because there is little emphasis on producing good documentation and models to support software creation and evolution of large, complex systems.

The claims made by agile process proponents and critics lead to questions about what practices, techniques, and infrastructures are suitable for software development in today's rapidly changing development environments. In particular, answers to questions related to the suitability of agile processes to particular application domains and development environments are often based on anecdotal accounts of experiences.

In this paper we present what we perceive as limitations of agile processes based on our analysis of published works on agile processes [14]. Processes that name themselves "agile" vary greatly in values, practices, and application domains. It is therefore difficult to assess agile processes in general and identify limitations that apply to all agile processes. Our analysis [14] is based on a study of assumptions underlying Extreme Programming (XP) [3,5,6,10], Scrum [12,13], Agile Unified Process [11], Agile Modeling [1] and the principles stated by the Agile Alliance. It is mainly an analytical study, supported by experiences on a few XP projects conducted by the authors.

## THE AGILE ALLIANCE
In recent years a number of processes claiming to be "agile" have been proposed in the literature. To avoid confusion over what it means for a process to be "agile", seventeen agile process methodologists came to an agreement on what "agility" means during a 2001 meeting where they discussed future trends in software development processes. One result of the meeting was the formation of the "Agile Alliance" and the publication of its manifesto (see http://www.agilealliance.org/principles.html). The manifesto of the "Agile Alliance" is a condensed definition of the values and goals of "Agile Software Development". This manifesto is detailed through a number of common principles for agile processes. The principles are listed below.

*1. "Our highest priority is to satisfy the customer through early and continuous delivery of valuable software."*
*2. "Business people and developers must work together daily throughout the project."*
*3. "Welcome changing requirements, even late in development."*
*4. "Deliver working software frequently."*
*5. "Working software is the primary measure of progress."*
*6. "Build projects around motivated individuals. Give them the environment and support they need, and trust them to get the job done."*
*7. "The best architectures, requirements, and designs emerge from self-organizing teams."*
*8. "The most efficient and effective method of conveying information to and within a development team is face-to-face conversation."*
*9. "Agile processes promote sustainable development."*
*10. "Continuous attention to technical excellence and good design enhances agility."*
*11. "Simplicity is essential."*
*12. "Project teams evaluate their effectiveness at regular intervals and adjust their behavior accordingly."*

## AN ANALYSIS OF AGILE PROCESSES

In this section we discuss the limitations of agile processes that we have identified, based on our analysis of the Agile Alliance principles and assumptions underlying agile processes. The next subsection lists the managerial and technical assumptions we identified in our study [14], and the following subsection discusses the limitations derived from the assumptions.

### Underlying Assumptions

The stated benefits of agile processes over traditional prescriptive processes are predicated on the validity of these assumptions. These assumptions are discussed in more details in another paper [14].

*Assumption 1:* Customers are co-located with the development team and are readily available when needed by developers. Furthermore, the reliance on face-to-face communication requires that developers be located in close proximity to each other.

*Assumption 2:* Documentation and software models do not play central roles in software development.

*Assumption 3:* Software requirements and the environment in which software is developed evolve as the software is being developed.

*Assumption 4:* Development processes that are dynamically adapted to changing project and product characteristics are more likely to produce high-quality products.

*Assumption 5:* Developers have the experience needed to define and adapt their processes appropriately. In other words, an organization can form teams consisting of bright, highly-experienced problem solvers capable of effectively evolving their processes while they are being executed.

*Assumption 6:* Project visibility can be achieved primarily through delivery of increments and a few metrics.

*Assumption 7:* Rigorous evaluation of software artifacts (products and processes) can be restricted to frequent informal reviews and code testing.

*Assumption 8:* Reusability and generality should not be goals of application-specific software development.

*Assumption 9:* Cost of change does not dramatically increase over time.

*Assumption 10:* Software can be developed in increments.

*Assumption 11:* There is no need to design for change because any change can be effectively handled by refactoring the code [9].

### Limitations of Agile Processes

The assumptions listed above do not hold for all software development environments in general, nor for all "agile" processes in particular. This should not be surprising; none of the agile processes is a silver bullet (despite the enthusiastic claims of some its proponents). In this part we describe some of the situations in which agile processes may generally not be applicable. It is possible that some agile processes fit these assumptions better, while others may be able to be extended to address the limitations discussed here. Such extensions can involve incorporating principles and practices often associated with more predictive development practices into agile processes.

*1. Limited support for distributed development environments:*

The emphasis on co-location in practices advocated by agile processes does not fit well with the drive by some industries to realize globally distributed software development environments. Development environments in which team members and customers are physically distributed may not be able to accommodate the face-to-face communication advocated by agile processes. In such cases, one can at least approximate face-to-face communication using technologies such as video-conferencing, but these technologies are expensive and not as effective as one would hope.

Face-to-face communication is as important in distributed environments as non-distributed environment, but it occurs less frequently and has to be planned in advance to ensure that all involved can participate. One can use such face-to-face meetings as major synchronization events in which geographically dispersed developers (1) are made aware of the progress made by others and (2) discuss plans for further evolving the product. In between such meetings, documentation (beyond code) becomes the primary form of communication. Good documentation of requirements and designs, produced and maintained in a timely manner, are essential to ensure that the distributed team members all maintain the same vision of the product to be built. This should not be interpreted as a requirement to document or model all aspects of software. Documentation and models should be created and maintained only if they provide value to the project and project stakeholders.

*2. Limited support for subcontracting:*

Outsourcing of software development tasks to subcontractors is often based on contracts that precisely stipulate what is required of the subcontractor. Subcontracted tasks have to be well-defined in the cases where subcontractors have to bid for the contract. In developing a bid a subcontractor will usually develop a plan that includes a process, with milestones and deliverables, in sufficient detail to determine a cost estimate. The process may be an iterative, incremental approach, but the subcontractor may have to make the process predictive by specifying the number of iterations and the deliverables of each iteration in order to compete.

It is possible that a contract can be written that allows a subcontractor some degree of flexibility in how they develop the product within time and cost constraints. This is certainly possible if the subcontractor has a good track record and can be trusted by the contracting company to develop a product that meets the contracting company's needs. A contract supporting agile development in the subcontractor environment should consist of two parts:

- Fixed Part: This part defines (1) the framework that constrains how the subcontractor will incorporate changes into the product (e.g., cost- and time-based criteria for accepting or rejecting changes to the Variable Part (see below) of the contract, (2) the activities that must be carried out by the subcontractor (e.g., quality assurance activities), and (3) requirements that are to be considered fixed and deliverables that must be delivered.
- Variable Part: This part defines the requirements and deliverables that can vary within the boundaries defined in the Fixed Part. This part can evolve within the constraints defined in the Fixed Part. At the time the contract is signed, a description of prioritized deliverables and requirements should be included.

*3. Limited support for building reusable artifacts:*

Agile processes such as Extreme Programming focus on building software products that solve a specific problem. Development in "Internet time" often precludes developing generalized solutions even when it is clear that this could yield long-term benefits. In such an environment, the development of generalized solutions and other forms of reusable software (e.g., design frameworks) is best tackled in projects that are primarily concerned with the development of reusable artifacts. This separation of the product-specific development environment from the reusable artifact development environment is a primary feature of the reuse-oriented framework called the *Experience Factory* developed by researchers at the University of Maryland at College Park [4]. The wide applicability of a reusable artifact requires that the process used to build the artifact emphasize quality control because the impact of low quality (in particular, severe errors) is as wide as the number of applications that reuse the artifact. On the other hand, timely development of reusable artifacts is desirable. While there seems to be a case for applying agile processes to the development of reusable artifacts, it is not clear how agile processes can be suitably adapted.

*4. Limited support for development involving large teams:*

Agile processes support process "management-in-the-small" in that the coordination, control, and communication mechanisms used are applicable to small to medium sized teams. With larger teams, the number of communication lines that have to be maintained can reduce the effectiveness of practices such as informal face-to-face communications and review meetings. Large teams require less agile approaches to tackle issues particular to "management-in-the-large". Traditional software engineering practices that emphasize documentation, change control and architecture-centric development are more applicable here. This is not to say that agile practices are not applicable in such environments. There may be opportunities for teams to use agile practices, but the degree of agility possible may be less than that found in smaller projects.

*5. Limited support for developing safety-critical software:*

Safety-critical software is software in which failure can result in direct injury to humans or cause severe economic damage. The quality control mechanisms supported by current agile processes (e.g., informal reviews, pair-programming) have not proven to be adequate to assure users that the product is safe. In fact there is some doubt that these techniques alone will be sufficient. Formal specification, rigorous test coverage, and other formal analysis and evaluation techniques included in software engineering approaches provide better, but also more expensive, mechanisms to tackle the development of safety- or business-critical software. Some agile practices can also bring benefits to the development of such software. For example, (1) test-first approaches requires one to define unit tests before writing code, (2) the early production of working code supported by the incremental, iterative process structure of agile processes supports exploratory development of critical software in which requirements are not well-defined, and (3) pair-programming can be an effective supplement to formal reviews. Therefore, it can be assumed that agile and formal software development are not incompatible, but can be combined when needed: Formal techniques may be used in an agile way to handle critical pieces of the software to increase quality and confidence.

*6. Limited support for developing large, complex software:*

The assumption that code refactoring removes the need to design for change may not hold for large complex systems in particular. In such software, there may be critical architectural aspects that are difficult to change because of the critical role they play in the core services offered by the system. In such cases, the cost of changing these aspects can be very high and therefore it pays to make extra efforts to anticipate such changes early. The reliance on code refactoring could also be problematic for such systems. The complexity and size of such software may make strict code refactoring costly and error-prone. Models can play an important role here, especially if tools exist for generating significant portions of the code from the models. This view of models as the central artifacts for evolving systems is at the heart of the Object Management Group's (OMG) Model-Driven Architecture (MDA) approach (see http://www.omg.org/mda).

There may also be systems in which functionality is so tightly coupled and integrated that it may not be possible to develop the software incrementally. In these cases an iterative approach in which code is produced in each iteration can still be used, but the code produced in each iteration will include all the pieces in various states of incompleteness.

## CONCLUSIONS

While it appears that there have been many software development project successes based on agile processes, so far most of these success stories are only anecdotal. Empirical data comparing the effectiveness and limitations of agile and non-agile approaches would greatly enhance our understanding of the true benefits and limitations of agile processes. In this paper we presented a list of limitations based on a study of principles and assumptions underlying a subset of the processes that claim to be "agile". Not all assumptions apply to all these processes. For example "Crystal Blue" a yet unpublished larger brother of "Crystal Clear" [7] will have good support for developing large software, but will probably be less "agile". It is clear, that certain domains are more amenable to agile development processes. Among them are Internet application domains, in which there are significant time-to-market pressure and the costs of upgrading to the next release are minimal. However, it is also clear that companies that develop long-lasting, large complex systems may not be able to use agile processes in their current form.

In general, some aspects of a software development project can benefit from an agile approach while others can benefit from a less-agile or more predictive approach. From this perspective, practical software development processes can be classified along a spectrum depending on their degree of "agility". At one extreme of the spectrum are the purely predictive processes in which the process steps are defined in detail early in the project, and project goals remain relatively stable throughout the execution of the process. At the other end of the spectrum are the purely agile processes in which process steps and project goals are dynamically determined based on analyses of (1) experiences gained with previously executed process steps, (2) similar experiences gained outside of the project, and on (3) changes in the requirements and development environment. From this perspective, the agility of a process is determined by the degree to which a project team can dynamically adapt the process based on changes in the environment and the collective experiences of the developers.

Practical processes lie somewhere in between the purely agile and purely predictive spectrum extremes. Current agile processes are close to the purely agile end of the spectrum, but they are not purely agile because they provide a process framework that constrains the form of processes that developers must follow. For example, most published works on agile processes stipulate an iterative, incremental process and advocate practices such as test-first code development, pair-programming, and daily review meetings with particular formats.

## ACKNOWLEDGEMENTS
Bernhard Rumpe's work was supported in part by the Bayerisches Staatsministerium für Wissenschaft, Forschung und Kunst through the Bavarian Habilitation Fellowship and the German Bundesministerium für Bildung und Forschung through the Virtual Softwaereengineering Competence Center (ViSEK). The work of Robert France and Dan Turk was supported in part by a grant from the Colorado Advanced Software Institute (CASI) and Qwest (CASI Project 5-30186).